\documentclass[conference]{IEEEtran}

\IEEEoverridecommandlockouts

\usepackage{alltt}

\usepackage{tikz}
\usepackage{upquote}
\usetikzlibrary{decorations.pathreplacing}

\usepackage{graphicx}
\usepackage{comment}
\usepackage{subfig}
\usepackage[normalem]{ulem} 
\usepackage{amstext}
\usepackage{pifont}
\usepackage{balance}
\usepackage{MathDefs}
\usepackage{docmute}
\usepackage{circuitikz}
\usepackage{amsmath,amsfonts,amssymb,amsthm}
\usepackage{hyperref}
\hypersetup{
  colorlinks   = true, 
  urlcolor     = blue, 
  linkcolor    = blue, 
  citecolor   = red 
}
\usepackage{nomencl}
\makenomenclature
\usepackage{color, soul} 

\usepackage{tcolorbox} 
\newtcolorbox{Educ}[1]{
 title=#1,
  beamer, 
  colback=xlightblue,
  colframe=blue!30,
  fonttitle=\bfseries,
  left=1mm,
  right=1mm,
  top=1mm,
  bottom=1mm,
  middle=1mm,
  breakable,
}

\usepackage{url}
\usepackage{chapterbib}

\newtheorem{lemma}{Lemma}

\usepackage{algorithm}
\usepackage{algpseudocode} 

\usepackage{tabularx}
\newcolumntype{L}[1]{>{\raggedright\let\newline\\\arraybackslash\hspace{0pt}}m{#1}}
\newcolumntype{C}[1]{>{\centering\let\newline\\\arraybackslash\hspace{0pt}}m{#1}}
\newcolumntype{R}[1]{>{\raggedleft\let\newline\\\arraybackslash\hspace{0pt}}m{#1}}

\def\BibTeX{{\rm B\kern-.05em{\sc i\kern-.025em b}\kern-.08em
    T\kern-.1667em\lower.7ex\hbox{E}\kern-.125emX}}
\begin{document}

\title{Performance Analysis of Empirical Open-Circuit Voltage Modeling in Lithium Ion Batteries, \\Part-1: Performance Measures}

\author{\IEEEauthorblockN{P. Pillai, J. Nguyen, and B. Balasingam}
}

\maketitle

\begin{abstract}
The open circuit voltage to the state of charge (OCV-SOC) characteristic is crucial for battery management systems. 
Using the OCV-SOC curve, the SOC and the battery capacity can be estimated in real-time. 
Accurate SOC and capacity information are important to carry out the majority of battery management functionalities that ensure a safe, efficient, and reliable battery pack power system. 
Numerous approaches have been reported in the literature for improved SOC estimation and battery capacity estimation. 
These approaches focus on various estimation and filtering techniques to reduce the effect of measurement noise and uncertainties due to hysteresis and relaxation effects. 
Even though all the existing approaches to SOC estimation rely on the OCV-SOC characterization, little attention was paid to investigating the possibility of errors in the OCV-SOC characterization and the effect of uncertainty in the OCV-SOC curve on SOC and capacity estimates.
In this paper, which is the first part of a series of three papers, the effect of OCV-SOC modeling error in the overall battery management system is discussed. 
The different sources of uncertainties in the OCV-SOC curve include cell-to-cell variation, temperature variation, aging drift, cycle rate effect, curve-fitting error, and measurement/estimation error.
The proposed uncertainty models can be incorporated into battery management systems to improve their safety, performance, and reliability. 
\end{abstract}



\section{Introduction}

Lithium-ion (Li-ion) batteries were first developed for use in consumer electronic devices, such as cell phones, laptops, and tablets for their energy density which was high enough to power consumer electronic devices for daily activities. 
Recently, Li-ion batteries have been widely adopted in various other applications such as power equipment, energy storage grids, and electric vehicles (EVs). 
Compared to other types of energy sources (e.g., hydrocarbons, hydrogen, etc.) the energy density of Li-ion batteries is extremely low. 
Despite that, the automotive industry is forecasted to adopt Li-ion batteries in transport electrification in an effort to eliminate energy sources that produce greenhouse gases \cite{sanguesa2021review}.

\begin{figure}[h!]
\begin{center}
\includegraphics[width=\columnwidth]{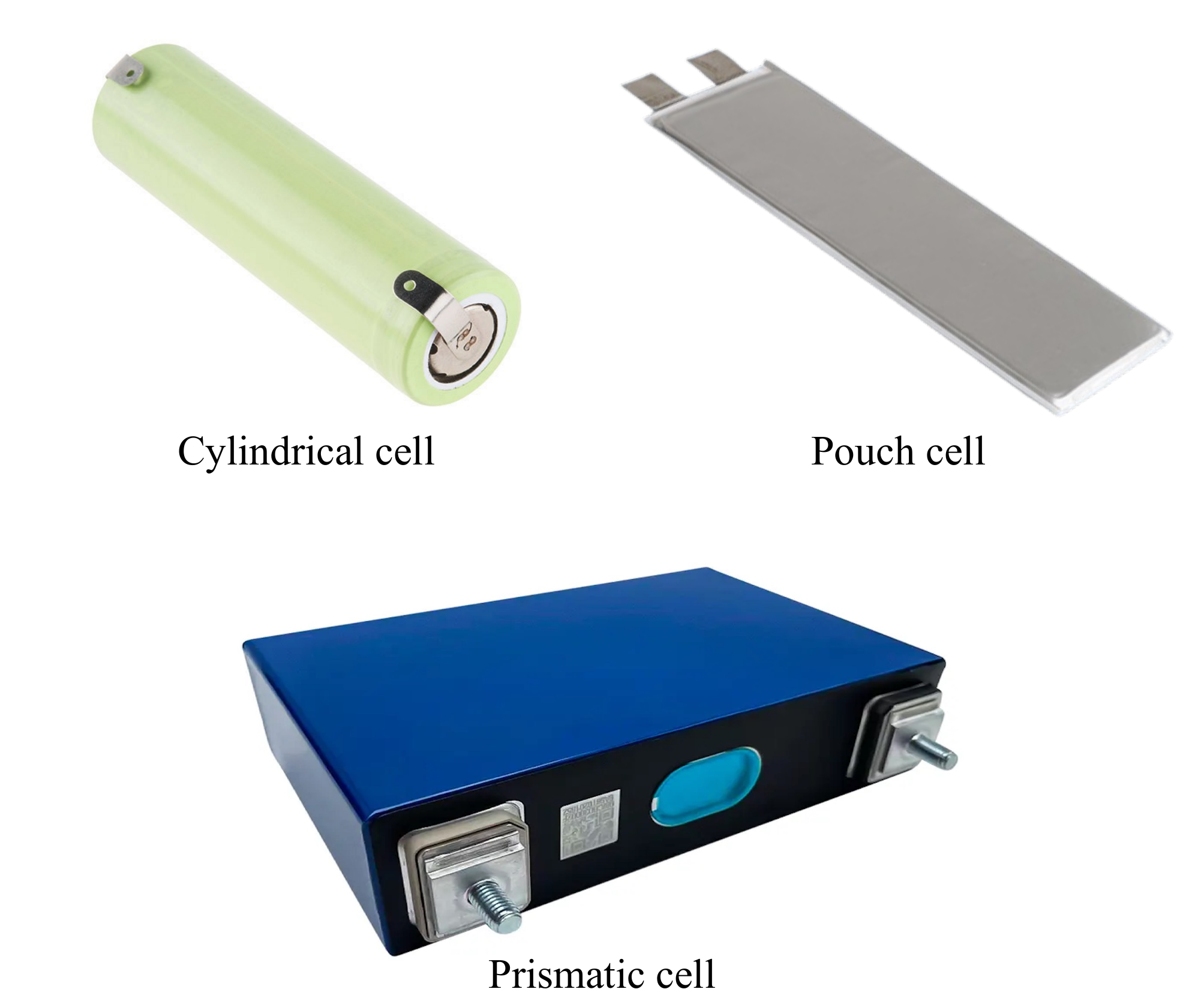}
\caption{Basic battery cells used in the EV industry.}
\label{fig:EVcells}
\end{center}
\end{figure}

One Li-ion battery cell provides a nominal voltage of about 3.8V. 
The amount of current over time (or energy) it can produce depends on the size of the electrodes and their composition. 
The material and electro-chemical research field is actively seeking ways to increase the energy density of basic Li-ion cells. 
For this, researchers experiment with various materials to improve the properties of the four primary components of a Li-ion rechargeable battery \cite{warner2015handbook}: anode, cathode, separator, and electrolyte.
The resulting high-energy density battery cell in present-day EVs can be in a pouch, prismatic, or cylindrical configurations. 
Figure \ref{fig:EVcells} shows the three types of basic cells used in commercial EV applications.  

The voltage requirement of EVs is in the range of 400V to 800V; this would require the series connection of more than 100 to 200 cells. 
To increase the capacity, hundreds of cells need to be connected in parallel.
In total, tens of thousands of cells need to be connected in series and parallel to make a battery pack that can power an EV.  
The Li-ion battery cells are made of combustible materials:
the metallic lithium is highly reactive with air, water, moisture, and steam \cite{chen2020thermal}.
Even though active lithium is substituted with lithium metal oxides in the cathode and lithium salts, the batteries can still be prone to combustion under high voltage and high-temperature conditions \cite{zhan2021promises}.
In addition, Li-ion batteries risk irreversible damages when frequently exposed to low voltage and over-discharge \cite{chen2021review}. 
Further, attempts to increase the energy density of Li-ion batteries \cite{lahiri2018building} have resulted in thinner electrodes that increase the likelihood of short circuits and thermal runaway in EV batteries. 
In order to ensure safety and reliability, each battery cell within a battery pack (of tens of thousands of cells) must be kept within specific voltage, temperature, and SOC conditions.
As a result, the battery management system (BMS) has become an integral component of a battery pack in high-power and EV applications.

A BMS performs various operations to ensure the safe, reliable, and efficient operation of a battery pack \cite{habib2023lithium}. 
Some of the important BMS functionalities are listed below:
\begin{enumerate}
\renewcommand{\labelenumi}{(\alph{enumi})}
\item 
Compute and report the status of the energy storage. 
The SOC indicator is the most basic functionality of a BMS.
In consumer electronic devices the SOC indicator serves as an approximate measure of the remaining time until loss of power. 
In high-power applications, SOC information is needed in many other BMS functionalities that are described in the remainder of this list. 

\item
Compute and report the time to shut down (TTS):
The TTS varies depending on the amount of load current. 
In consumer electronic applications, tasks that demand high CPU and memory involvement (e.g., video processing) require high current compared to other applications; the remaining TTS will be lower during video playing compared to a phone call. 
In EVs, highway driving requires relatively more current compared to city driving \cite{keil2015aging}; the remaining TTS (or remaining range) information is more critical for EVs compared to consumer electronic devices.
The TTS is computed based on SOC, battery capacity, and internal resistance \cite{pillai2022approach}.

\item
Charging:
A battery pack (or module) is charged using a constant current or constant voltage protocol \cite{ghaeminezhad2022charging}. 
During charging, the voltage across each cell increases. 
The BMS monitors the voltage at every level of a battery pack; 
the charger needs to be disconnected whenever the voltage across a particular cell or module exceeds the safety threshold. 
The SOC, capacity, and resistance information is used by advanced charging algorithms to develop state of health (SOH) aware charging strategies \cite{abdollahi2017optimal}. 

\item
Charge balancing \cite{habib2022energy}:
Cell imbalance causes an uneven rise in voltages among cells; the weakest cell sees its voltage reach the safety threshold first, thereby preventing the remaining cells from being fully charged. 
Similarly, the weakest cell causes the battery to prematurely shut down during discharging. 
One of the important tasks of a BMS in multi-cell battery packs is to use cell-balancing strategies to prevent the battery pack from premature shutdown during charging and discharging \cite{hoekstra2022optimal}. 
Basic cell balancing can be done only based on voltage measurements; advanced cell balancing strategies will require SOC, battery capacity, and resistance information.

\item
Thermal balancing \cite{mallick2023thermal}:
Li-ion battery packs perform best at room temperature; 
the available power decreases at lower temperatures and the risk of thermal runaway increases at higher temperatures. 
Heating and cooling mechanisms are integral parts of a BMS \cite{wu2019critical}. 
Advanced BMS algorithms use SOC and internal resistance estimates to effectively regulate the temperature within a battery pack \cite{kumar2022model}. 
 
\item 
Remaining useful life estimation:
Battery resistance increases with its usage and causes the available power to decrease \cite{chang2022prognostics}. 
Power fade (PF) is one of the indicators of the state of health. 
Similarly, capacity fade (CF) is another indicator of the state of health of the battery. 
Advanced BMSs contain state-of-health models to predict the remaining useful life of a battery \cite{vennam2022survey}. 

\end{enumerate}

A majority of BMS functionalities described above require three crucial parameters of a battery: 
\begin{itemize}
\item 
The SOC
\item
Battery capacity, and
\item
Internal resistance
\end{itemize}

The SOC is defined as the ratio of the remaining Coulombs to the battery capacity as follows: 
\begin{align}
{\rm SOC} = \frac{\rm Remaining \,\,Coulombs \,\, (Ah)}{\rm Battery \,\, Capacity \,\, (Ah)}
\label{eq:SOCratio}
\end{align}
It is also common to define the SOC in percentages as follows: 
\begin{align}
{\rm SOC} = \frac{\rm Remaining \,\,Coulombs \,\, (Ah)}{\rm Battery \,\, Capacity \,\, (Ah)} \times 100 
\label{eq:SOCperc}
\end{align}
In this paper, the SOC is referred to as the ratio \eqref{eq:SOCratio} in all equations that involve the computation of OCV from SOC and vice versa. 
The percentage definition of \eqref{eq:SOCperc} is used in graphs and displays. 

The Coulomb counting approach can be used to estimate the SOC as follows 
\begin{align}
\begin{aligned}
s(k) &= s(k-1) + \frac{1 }{Q}  \int_{t_{k-1}}^{t_k} i(t) dt  \\
& \approx s(k-1)  + \frac{\Delta_k i(k)}{Q}
\end{aligned}
\end{align}
where
$s(k)$ denotes the SOC at a certain time instance $t_k$ and
$Q$ denotes the battery capacity. 
However, the Coulomb counting approach suffers from several deficiencies \cite{Movassagh2021}. 
Improvements to the Coulomb counting approach rely on accurate OCV-SOC characterization. 

The SOC means little without the knowledge of battery capacity which fades over time due to the loss of active material within the battery cell \cite{liu2022capacity}. 
The capacity fade occurs due to environmental factors (extreme temperature) and usage factors (over-charge and over-discharge).  
The BMS plays a dual role when it comes to battery capacity:
it utilizes algorithms to accurately estimate battery capacity and employs various control measures, such as battery thermal management and cell balancing, to reduce capacity fade.  

The internal resistance changes with temperature, age, and, to some extent, the battery SOC \cite{wu2023battery}. 
Compared to the battery capacity and the OCV-SOC curve, the change in the internal resistance due to temperature and age is very significant. 
On the other hand, unlike battery capacity and the OCV-SOC curve, the internal resistance can be estimated relatively quickly by employing model-based signal processing techniques \cite{pillai2022optimizing,hou2022review}. 

The focus of this paper is on the OCV-SOC curve which is crucial for the accurate estimation of both the SOC and the battery capacity. 
For the first time, a mathematical relationship is established between the uncertainty in the OCV-SOC model to the uncertainty of the SOC and battery capacity estimates. 
The SOC estimation error depends on two factors:
the shape of the OCV-SOC curve and its uncertainty. 
While the shape of the OCV-SOC curve is based on the battery chemistry and can not be controlled by the BMS, the uncertainty of the OCV-SOC model can be reduced by incorporating better OCV modeling approaches. 
This paper identifies five different sources of uncertainties that can be quantified through empirical modeling approaches.

The remainder of this paper is organized as follows. 
Section \ref{sec:OCVcurve} presents the OCV-SOC representation through electrochemical interpretation. 
Existing standards about empirical OCV-SOC characterization are summarized in Section \ref{sec:EmpiricalOCV}. 
Section \ref{sec:OCVmodelingLiterature} presents a literature review of recent empirical OCV modeling approaches. 
Section \ref{sec:OCVerrorEffects} presents the mathematical derivation of the effect of the uncertainty of the OCV model in two important quantities needed for BMSs: the SOC and the battery capacity. 
In Section \ref{sec:OCVerrorModels}, various sources of errors that contribute to uncertainties in the OCV model of a BMS are presented;
models are presented to empirically quantify these 
Finally, Section \ref{sec:conc} concludes the paper.

\section{The OCV-SOC Curve}
\label{sec:OCVcurve}

\begin{figure}[h]
\begin{center}
\includegraphics[width=\columnwidth]{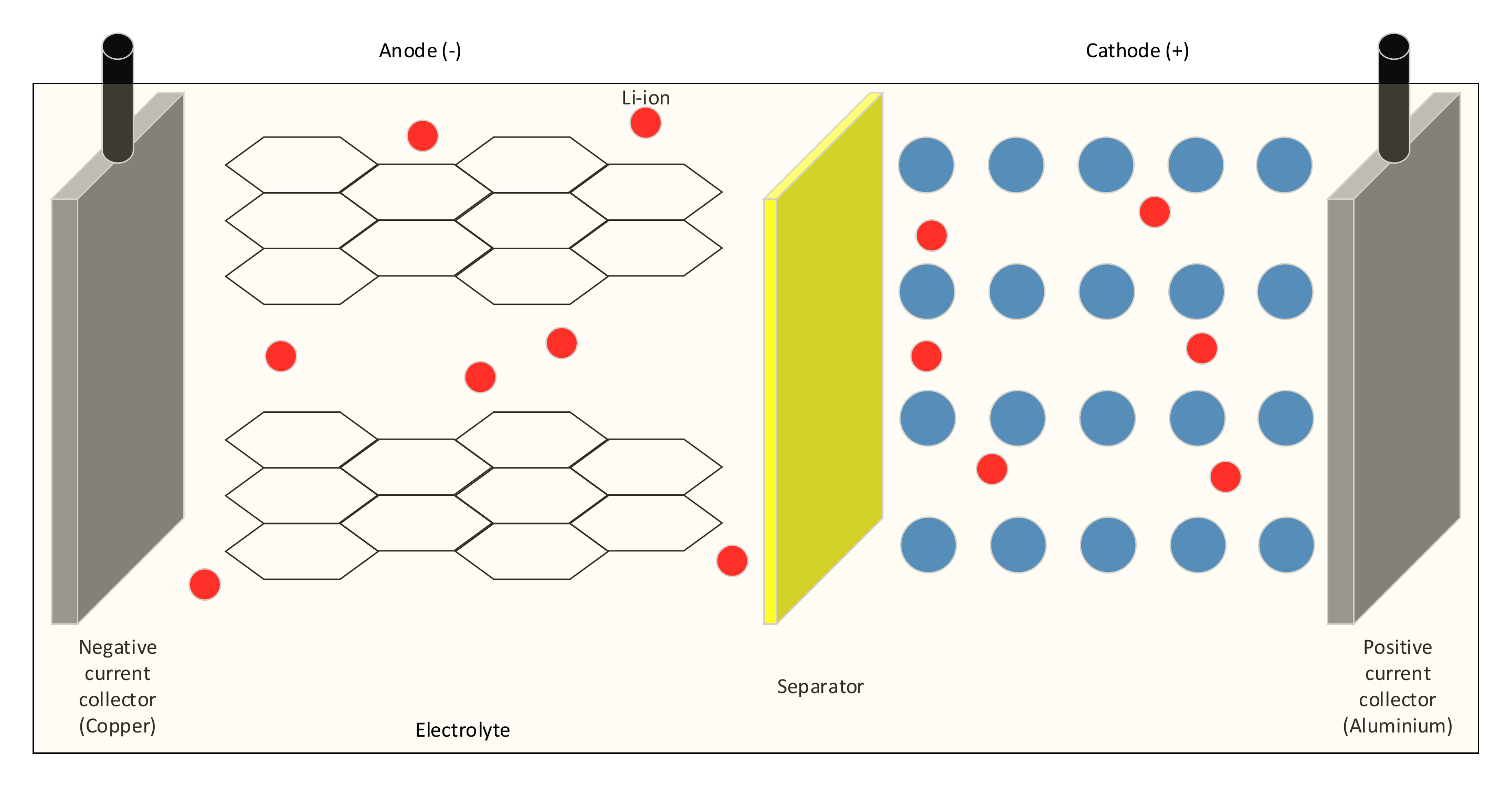}
\caption{Diagram of a battery cell \cite{balasingam2023robust}}
\label{fig:li-ionCell}
\end{center}
\end{figure}

Figure \ref{fig:li-ionCell} shows a generic diagram of a Li-ion battery cell. 
When a charging voltage is applied to the battery, it induces the movement of Li-ions from the positive electrode (cathode) to the negative electrode (anode); due to this, the chemical equilibrium between the anode and cathode is disrupted. 
When the charger is stopped, it takes time for the chemical equilibrium to be regained. 
The time to regain the chemical equilibrium is known as relaxation time and is modeled using Resistive-Capacitive (RC) elements in the electrical equivalent circuit models \cite{pillai2022optimizing}. 
During discharging, the direction of movement of Li-ions is reversed; the relaxation mechanism after the discharging process is very similar to that after charging and is modeled using the same RC elements in a BMS.

When the battery is relaxed (after charging or discharging) the chemical potential at equilibrium is written using the Nernst equation as \cite{birkl2015parametric}
\begin{align}
\mu = E_0 - KT \ln \left[ \frac{x}{1-x} \right]
\label{eq:Nernst}
\end{align}
where $E_0$ is the standard redox potential, 
$K$ is the Boltzmann constant,
$T$ is the temperature, and 
$x$ is the ratio of intercalated sites to available sites in the host structure (anode). 
The above equilibrium happens on a regional basis and there can be several such regions in a single cell. 
The electrochemical theory allows one to precisely derive the chemical potential given precise knowledge of material properties. 
It also relates the chemical potential to the electric potential between battery cell terminals which is measured in volts (V). 

Electrochemical modeling of OCV requires precise knowledge of the material composition of the electrodes and their properties. 
The availability of such knowledge alone does not guarantee an accurate OCV model because of possible material transformation due to various environmental, usage, and aging factors. 
State-of-the-art battery management systems rely on empirical modeling to obtain the OCV-SOC characteristics.

In practical BMSs, the OCV-SOC characterization is done through empirical methods where 
the OCV-SOC characterization data (pair of OCV, SOC values) is obtained from sample batteries. 
This data is either stored as a table or fitted to a mathematical model through curve-fitting methods;
the procedure is known as OCV characterization or OCV-SOC characterization \cite{pattipati2014open,pillai2022open}. 
The OCV-SOC characterization process is standardized for different industry applications. 
Section \ref{sec:EmpiricalOCV} reviews existing standards for empirical OCV characterization. 
Section \ref{sec:OCVmodelingLiterature} discusses ongoing research papers that propose improved approaches to OCV-SOC characterization.

\section{Empirical OCV Characterization Standards}
\label{sec:EmpiricalOCV}

Empirical approaches to OCV parameterization can be classified into the following two categories: Galvanostatic Intermittent Titration Technique (GITT) and the low-rate cycling method. These two approaches are briefly discussed in the remainder of this section.

\subsection{Galvanostatic Intermittent Titration Technique (GITT)}
\label{sec:GITT}

In this approach, individual OCV-SOC points are measured intermittently and recorded in a way that the resulting data spans the entire SOC range. 
Here, the SOC is changed by applying a constant current called the HPPC (Hybrid Pulse Power Characterization) current calculated from the discharge power and battery size factor provided by the manufacturer. Two standards for the level of HPPC current are mentioned in the literature: 
one at low current and the other at high current \cite{christophersen2015battery}. 
Existing standards stipulate discharging the cell in 10\% steps and applying a 10-second charge/discharge pulse at each step to estimate other battery parameters such as the resistance and RC components. 
The rest of 1 hour is standardized for allowing the battery to achieve cell equilibrium potential. Just before the next HPPC discharge step, the OCV is measured.
  
The resulting OCV-SOC characterization from this standard will consist of 11 OCV measurements corresponding to the following 11 SOC values: ($0\%,$ $10\%,$ $\ldots,$ $100\%$). Smoothing or interpolation techniques were employed to develop the OCV-SOC curve between these SOC values. 
Modifications to the GITT approach were done later with more OCV measurement points and lower discharge currents.
The following are some of the considerations that make the GITT technique undesirable for OCV modeling: (1) the lower number of OCV-SOC points and resulting inaccuracy, (2) the inability of certain battery chemistries and aged batteries to reach equilibrium potential even after longer rest periods, and (3) the long duration of the test.

\subsection{Low-rate cycling method}
\label{sec:lowrateOCV}

In this approach (see \cite{pillai2022open,pattipati2014open} for a review), a battery is discharged and then charged using the same low C-Rate while continuously collecting the voltage and current data. Owing to the low current rate, the low-rate cycling method is also called the Coulomb titration (CT) technique.
The time taken to fully discharge and charge the battery can be used to compute the discharge capacity and charge capacity, respectively. 
Based on the computed capacities, the SOC values corresponding to each time instant in the collected data will be computed. 
The OCV (corresponding to a certain SOC) is then computed to be the average of the charge and discharge voltages.  
Curve fitting techniques are also used to capture the OCV-SOC values in a few OCV model parameters. 
Unlike the GITT method, there is no known standardized rate to cycle the battery in this technique. Usually, a C/25 rate or a lower rate is adopted \cite{barai2019comparison}.

In both of the experimental OCV characterization approaches described above (the GITT method and low-rate cycling method) the knowledge of battery capacity is needed. 
In the GITT method, the battery capacity information is needed to select the load-current and time duration such that the SOC is changed by $10\%$ each time.
In the low-rate cycling method \cite{pillai2022open,pattipati2014open}, the capacity information is used to compute the SOC which is then modeled against OCV. 
For the GITT method, the capacity needs to be estimated using a slow discharge test \cite{barai2019comparison}. 
For the low-rate cycling approach, the battery capacity can be estimated using the same low-rate cycle data collected for the OCV characterization. 
Some existing works also proposed to use label capacity for OCV modeling \cite{liu2023capacity}; however, the available capacity of a battery can be different from its label value.  
It was shown in \cite{pattipati2014open} that the rated capacity, computed at the same C-Rate as the OCV test yielded a consistent OCV model at multiple temperatures.

\section{Ongoing Research About OCV Modeling}
\label{sec:OCVmodelingLiterature}

In the past few decades, significant research has been done on Li-ion batteries and their constituted modeling, both industrially and academically. This section provides a comprehensive review of research articles particularly investigating errors in the OCV modeling of a battery. The review is presented in five subsections---the first four subsections are indicative of the different uncertainties investigated previously in OCV modeling (the cell-to-cell variation, temperature variation, aging drift, and hysteresis effect) and the final section summarizes the exploration of online OCV modeling approaches.

\subsection{Modeling the Effect of  Cell-to-Cell Variations on OCV}

A series connection of around 100 to 200 cells is needed to meet the voltage requirement in EVs. To properly maintain and balance these cells, a fundamental knowledge of the OCV-SOC relationship and capacity of each individual cell is essential. 
Several works have reported that there is variability observed in the performance of individual cells in a pack \cite{dubarry2010origins}. Thus, an OCV-SOC curve that was computed from a sample battery cell does not perfectly align with the remaining cells it is intended to represent. This change in battery characteristics is accelerated under varying operating conditions, temperature and age \cite{zilberman2020simulation}. Recent works have focused on analyzing the effect of inaccurate gradients and hysteresis effects on the OCV-SOC relationship in Li-ion batteries \cite{farmann2017study}. Although the effects of cell-to-cell variations are investigated, most empirical and averaging approaches are developed on the basis of single cells, and may not apply in multi-cell configurations \cite{truchot2014state}. Thus, more research is needed to consider the cell-to-cell variations in OCV modeling of new and aged batteries.

\subsection{Modeling the Effect of Temperature on OCV}
\label{sec:OCV-Temp-Literature}

The operating conditions of EV batteries are not predictable and often have varying temperature ranges. 
Increasing attention is being given to modeling the effect of temperature for developing an accurate OCV-SOC model. Recent works observe the error in the OCV-SOC model due to varying temperatures and quantified the root mean square values in SOC error \cite{hemi2018open, knap2021effects}. 
In \cite{wang2020new}, a temperature model was defined using five coefficients and the OCV-SOC mapping error at temperature ranges -20 to 40$^{\circ}$ was studied. The results from this study showed that the OCV-SOC mapping error was reduced by 13.3\% when a temperature model was incorporated. 
Future work is thus required to study the dependence on temperature and develop a model capable of reducing the error in the OCV-SOC model, specifically for use in Li-ion batteries in EV applications.

\subsection{OCV Modeling of Aging Effects}

Characterization of a cell's OCV-SOC curve is done using parameters that are derived in laboratory settings for a sample battery. The aging of a battery has multifaceted effects on its performance, the prominent one being the increase in impedance due to the electrochemical phenomenon of SEI formation \cite{iurilli2021use}. Capacity fade is another effect of aging that occurs due to the loss of active material. Ongoing work aims to develop models for depicting the effect of aging while characterizing the OCV-SOC relationship of the battery. In \cite{lavigne2016lithium}, an OCV curve with a one-parameter variation for aging is developed and tested on four lithium batteries. The use of an aging model is based on the analysis of cell equilibrium voltage between two points in the OCV-SOC curve. Thus, further modeling of aging due to other electrochemical phenomena, model diagnosis and evaluation \cite{che2023health} is required for accurate OCV-SOC characterization.

\subsection{Online OCV Modeling}

OCV-SOC modeling is an offline process during which the data spanning the entire SOC region needs to be collected. However, some works in the literature presented approaches to estimate the OCV-SOC parameters in real-time \cite{meng2020comparative, cui2022online}. Similarly, online estimation approaches for battery degradation modeling have also been presented in the literature \cite{park2022novel}. It is expected that the uncertainty associated with the online OCV parameter estimation is significantly high compared to offline approaches.
Online data-driven approaches need confident estimates of battery parameters and without accurate modeling of the OCV-SOC curve of the battery, these online approaches are likely to perform poorly in SOC estimation and consequently in BMS.

\subsection{Accounting for Hysteresis Effects in OCV Modeling}

The hysteresis phenomenon in a battery is defined as the difference in cell equilibrium voltage observed at the same SOC, varying due to charge and discharge conditions. 
Several causes of hysteresis voltage are investigated in literature and recent battery modeling research has been focused on developing robust hysteresis models for application in OCV-SOC modeling \cite{jiao2022local, huang2023battery}. 
A simplistic approach is to average the voltages between the charge and discharge cycles (pseudo-OCV modeling) as the hysteresis voltage was assumed to be equal and opposite in both cycles \cite{barai2019comparison}. 
However, recent works have found the existence of an asymmetric effect of hysteresis voltage on the OCV \cite{yu2022study}. This asymmetric behavior is attributed to the varying availability of graphite on the electrode during charging and discharging. Further, the movement of graphite particles is also strongly influenced by factors such as SOC, temperature, current rate, aging, stress and vibration shock \cite{huang2023battery}. Thus, more research is required to develop and evaluate hysteresis models for accurate OCV-SOC modeling.

\section{Quantitative Analysis of OCV Modeling Error}
\label{sec:OCVerrorEffects}

The previous two sections summarized the existing standards of OCV-SOC modeling and the ongoing research efforts to improve its accuracy. 
In this section, specific insights are provided to qualitatively understand the effect of OCV modeling errors in the functionalities of a BMS.

\subsection{SOC Estimation}
\label{sec:SOClookup}

\begin{figure}[h!]
\begin{center}
\includegraphics[width = \columnwidth]{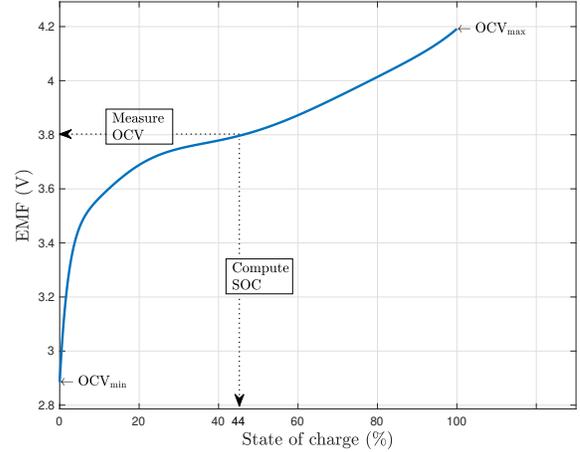}
\caption{SOC estimation based on OCV.} 
\label{fig:SOCest}
\end{center}
\end{figure}

Figure \ref{fig:SOCest} describes the SOC estimation approach based on the OCV-SOC curve. 
Here, the OCV-SOC curve is assumed to be stored in the form of a function \cite{pillai2022open}, for example, the Nernst OCV-SOC model is given as 
\begin{align}
{\rm OCV} = f_{\rm OCV} (s) =& k_0  + k_1 \ln(s) + k_2 \ln(1-s) 
\label{eq:nernst}
\end{align}
where $s$ denotes the SOC. 
The Nernst model in \eqref{eq:nernst} is based on Nernst electrochemical representation \eqref{eq:Nernst}.
Instead of using the physics-based parameters used in \eqref{eq:Nernst}, the empirical approaches seek to estimate OCV model parameters $k_0, k_1$ and $k_2$ based on data. 
This way, the BMS developers don't need to look for exact information on the chemical compositions of the battery components.
Empirical modeling approaches also add additional terms to improve the fitting accuracy of the observed data;
a list of such models can be found in \cite{pillai2022open}.

Assuming that the parameters $k_0, k_1, k_2$ of the OCV-SOC model are known, the SOC can be computed for a measured OCV as follows
\begin{align}
s  = f^{-1}_{\rm OCV} ({\rm OCV})
\label{eq:ocv_inv}
\end{align}
where $f^{-1}(\cdot)$ denotes the inverse function. 
When the function is not invertible in closed form, numerical root-finding techniques are employed to find SOC for OCV that can be measured after relaxing the battery. 
Alternatively, equivalent circuit model approaches are used to estimate the OCV of an active battery for real-time SOC estimation approaches. 

Let us denote the measured (or estimated) OCV value as follows 
\begin{align}
\hat E = E + \tilde E
\label{eq:Ehat}
\end{align}
where $E$ denotes the true OCV,
$\hat E$ denotes the OCV that is used for SOC estimation, and
$ \tilde E$ denotes the uncertainty in the OCV. 
Here, an estimate $\hat E$ of the OCV could be obtained by measuring it after resting the battery or by employing ECM-based estimation algorithms \cite{pillai2022optimizing}.

In this paper, the OCV uncertainty $ \tilde E$ is assumed to be a zero-mean Gaussian variable, i.e., 
\begin{align}
\tilde E \sim \cN (0, \sigma_\rE^2 )
\label{eq:Etilde}
\end{align}
where $\cN$ denotes the normal distribution and
$\sigma_\rE$ denotes the standard deviation of OCV uncertainty.
The uncertainty of OCV could come from various sources such as
cell-to-cell variation, hysteresis effect, relaxation effect, and the error introduced by various estimation algorithms utilized to estimate hysteresis and relaxation voltages.  
Refer to Section \ref{sec:OCVerrorModels} for formal definitions of various sources of errors that contribute to the OCV error in \eqref{eq:Ehat}. 

For a given OCV estimate, $\hat E$, the SOC is determined by the voltage lookup as in (\ref{eq:ocv_inv}) (i.e)
\begin{align}
\hat s  = f^{-1} ( \hat E)
\label{eq:hat-s}
\end{align}
where $f^{-1}_{\rm OCV}$ from \eqref{eq:ocv_inv} is denoted by $f^{-1}$ for ease of notation.

Now, the characteristics of the SOC estimate \eqref{eq:hat-s} need to be derived. 
The following lemma derives those quantities. 

\begin{lemma}[Mean and variance of the OCV based SOC estimate]
\label{lemma-hat-s}
\begin{align}
\mathbb{E}(\hat s) &= \bbE (f^{-1} ( \hat E)) \approx s \label{eq:E(hat-s)} \\  
\mathbb{E}((\hat s - s)^2) & = \sigma_\rs^2(s) \approx  \left (\frac{1}{f'(s)}\right)^2 \sigma_\rE^2 \label{eq:SOCerrorVar}
\end{align}
where $\bbE(\cdot)$ denotes the expectation operator. 
\end{lemma}

The mean and variance of the SOC estimate can be derived by expanding $f^{-1}(\hat E)$ about the true OCV, $E$, using Taylor's series expansion.
\begin{proof}
See \cite{sundaresan2022fast}.
\end{proof}

\begin{figure}[h]
\begin{center}
\includegraphics[width = \columnwidth]{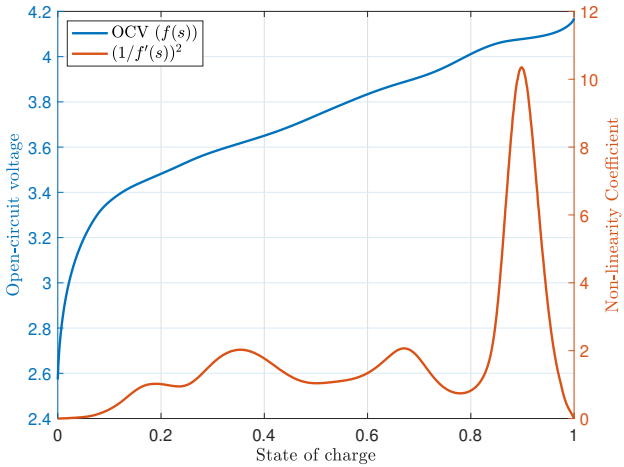}
\caption{SOC estimation based on OCV.} 
\label{fig:CapEstErrorCoefficient}
\end{center}
\end{figure}

The variance of the SOC estimation error, defined in \eqref{eq:SOCerrorVar} is shown to be a function of two coefficients;
the first one $({1}/{f'(s)})^2$ is denoted as the non-linearity coefficient and the second one $ \sigma_\rE^2$ is denoted as the OCV uncertainty coefficient. 
Figure \ref{fig:CapEstErrorCoefficient} shows computed values of the non-linearity coefficient for a particular OCV-SOC curve; it can be noticed that whenever the OCV curve is less steep (relatively flat) against SOC, the non-linearity coefficient increases; this indicates the fact that SOC estimation error is high when the OCV-SOC curve is relatively flat.
Conversely, when the OCV steeply increases against SOC (especially at low SOC regions) the non-linearity coefficient is very low. 
Section \ref{sec:OCVerrorModels} is dedicated to defining various factors influencing the OCV uncertainty coefficient $ \sigma_\rE^2$. 

\subsection{Battery Capacity Estimation}
\label{sec:CapacityEstimation}

\begin{figure}[h]
\begin{center}
\includegraphics[width = \columnwidth]{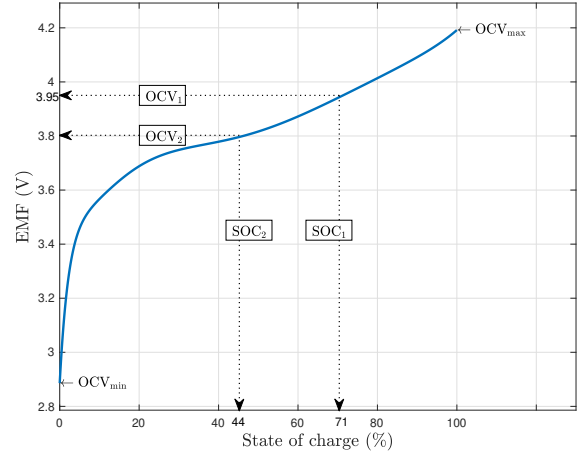}
\caption{OCV-based capacity estimation.} 
\label{fig:cap_est}
\end{center}
\end{figure}

The open circuit voltage model of a battery can be exploited to estimate the battery capacity.
Consider the scenario illustrated in Figure \ref{fig:cap_est} where the battery measured ${\rm OCV = OCV_1}$ at the start of the experiment. 
The battery is then discharged by extracting $C$ Coulombs (measured in Ah) from it.
At the end of this discharge, and after sufficiently resting the battery, the battery measured ${\rm OCV = OCV_2}$. 

Now, for the two OCV measurements in Figure \ref{fig:cap_est}, the corresponding SOC values can be obtained as 
\begin{align}
{\rm SOC_1}  & = f_{\rm OCV}^{-1}( {\rm OCV_1} )  \label{eq:SOC_1}\\
{\rm SOC_2}  & = f_{\rm OCV}^{-1}( {\rm OCV_2} ) \label{eq:SOC_2}
 \end{align}
The change in SOC is equal to the change in Coulombs normalized by the battery capacity, i.e., 
\begin{align}
{\rm d_{SOC}} = {\rm SOC_2}-{\rm SOC_1} = \frac{C}{Q}
\label{eq:dSOC}
\end{align}
where 
$C$ denotes the change in Coulombs and 
$Q$ denotes the battery capacity. 
Using the relationship \eqref{eq:dSOC}, the inverse battery capacity can be estimated as follows 
\begin{align}
\hat Q_\ri  = \frac{\rm d_{SOC}}{C}
\label{eq:Qest}
\end{align}
where $Q_\ri $ denotes the inverse of the battery capacity. 
Battery capacity is better estimated in inverse form; for more details, the reader is referred to \cite{sundaresan2022fast,bfg_part1}.

For capacity estimation \eqref{eq:Qest}, at least two estimates of SOC, as denoted in \eqref{eq:SOC_1} and \eqref{eq:SOC_2}, are needed. 
Let us denote the corresponding variances of the estimation errors as $\sigma^2_s(1)$ and $\sigma^2_s(2)$, respectively. 
Then, the capacity estimation in \eqref{eq:Qest} can be shown to have zero-mean and the following variance of the estimation error 
\begin{align}
R_Q = \frac{\sigma^2_s(1) +  \sigma^2_s(2)}{C^2 \hat Q_i^4}
\end{align}
where 
$\sigma^2_s(1)$ and $\sigma^2_s(2)$ are defined in \eqref{eq:SOCerrorVar}.

\section{Performance Measures of OCV Models}
\label{sec:OCVerrorModels}

Having discussed the importance of accurate OCV-SOC modeling in battery management systems, this section proposes some qualitative metrics by which an OCV-SOC model can be measured.

\subsection{Cell-to-Cell Variation}
\label{sec:C2Cvar}

In typical battery management systems, the OCV parameters are estimated from sample cells that are to be used in a particular application. 
The battery manufacturing process is very precise and the cells made through the same process are nearly identical. 
However, cell-to-cell variations are observed in brand-new commercial battery cells \cite{pillaiITEC2023}. 
In addition to manufacturing differences, cell-to-cell variations can also be caused by varied exposure to environmental and usage conditions during aging.  
The remaining variance can be modeled as 
\begin{align}
\tilde E_{\rm c2c}(s) \sim \cN (\mu_{\rm c2c}(s), \sigma_{\rm c2c}^2 (s) )
\end{align}
where $\tilde E_{\rm c2c}(s)$ denotes the modeling error in OCV at a certain SOC $s \in [0,1].$
It is hypothesized that the above modeling error due to cell variations is Gaussian distributed with mean $\mu_{\rm c2c}$ and s.d. $\sigma_{\rm c2c}^2 (s)$.
There are no existing works to confirm or disprove the above hypothesis. 
Some preliminary works to be reported in \cite{slowOCVp3} computes $\sigma_{\rm c2c}^2 (s) $ using empirical observations. 

\subsection{Temperature Variation}

The variations of the OCV (with respect to a certain SOC) at different temperatures are widely reported in the literature. 
Some results presented in \cite{pattipati2014open} showed that the temperature differences can be reduced by the normalized OCV modeling approach. Despite that, temperature changes in the battery will cause the assumed OCV model to be different from the ground truth.
The following model is proposed to absorb the variance due to temperature 
\begin{align}
\tilde E_{\rm T}(s) \sim \cN (\mu_{\rm T}(s), \sigma_{\rm T}^2 (s) )
\end{align}
where $\tilde E_{\rm T}(s)$ denotes the modeling error in OCV at a certain SOC $s \in [0,1]$ due to the temperature.

To reduce variations due to temperature, the OCV-SOC characterization could be repeated at several temperatures and the resulting temperature-dependent curves could be stored (Section \ref{sec:OCV-Temp-Literature} provides a literature review of such approaches); this results in increased complexity of the BMS.
It is also important to note that the differences between battery core temperature and the surface temperature \cite{kumar2022model} may affect the performance of temperature-dependent OCV-SOC models in practical BMSs.

\subsection{Aging Drift}

Similar to temperature, the variance of the OCV-SOC model due to aging is proposed to be modeled as a zero mean Gaussian random variable, i.e.,
\begin{align}
\tilde E_{\rA}(s) \sim \cN (\mu_{\rm A}(s), \sigma_{\rm A}^2 (s) )
\end{align}
where 
$\tilde E_{\rA}(s)$ denotes the difference between the true OCV model and the assumed OCV model due to aging,  
$\mu_{\rm A}$ and $ \sigma_{\rm A}^2 (s)$ denote the corresponding mean and variance of the OCV at a certain SOC $s \in [0,1]$ due to aging, respectively. 

The variations of the OCV due to aging can be reduced by performing the OCV characterization on artificially aged cells and by storing the corresponding parameters for different levels of aging. 
However, the real-world aging of a certain EV battery depends on numerous external factors that cannot be replicated in a laboratory setting.

\subsection{Cycle-Rate Error}
\label{sec:CrateError}

An OCV characterization test is usually done at $C/N$ rate, where $N$ is used to define the $C/N$ rate of constant current used to perform the data collection. 
The voltage drop across the resistance is written as
\begin{align}
\text{voltage drop} = \text{current} \times \text{resistance} 
\end{align}
Ideally, this voltage drop should be as low as possible for reaching the complete range of battery operation. As the resistance is a quantity that cannot be controlled or reduced, the current rate is manipulated to reduce this voltage drop.
Further, the selected current rate involves a trade-off between reaching the extreme values of SOC and the experimental time. A higher C-Rate takes only a few hours for the OCV experiment but the voltage drop across the resistor increases and vice versa.
For a battery with high internal resistance, a lower-current rate is required for a better relative range of operation of the battery. Hence, accurate modeling of the OCV-SOC curve is affected by the current during the OCV characterization test and it is hypothesized that the modeling error due to cycle rate is Gaussian distributed with mean $\mu_{\rm crate}$ and s.d. $\sigma_{\rm crate}^2 (s)$.
The variance due to cycle rate error can be modeled as 
\begin{align}
\tilde E_{\rm crate}(s) \sim \cN (\mu_{\rm crate}(s), \sigma_{\rm crate}^2 (s) )
\end{align}
where $\tilde E_{\rm crate}(s)$ denotes the modeling error in OCV at a certain SOC $s \in [0,1].$

Previous works in OCV characterization do not have a standard cycle rate and usually, a $C/25$ or $C/32$ rate was adopted. No existing works define the theoretical basis for selecting a particular rate and a lack of analysis on its effect on OCV error is noticed. The effect of C-Rate on the OCV-SOC curve is confirmed from the analysis of modeling using different C-Rates in \cite{slowOCVp3}.

\subsection{Curve Fitting Error}
\label{sec:MeasEstError}

The OCV-SOC data obtained through the low-rate cycling approach (see details in Section \ref{sec:lowrateOCV}) will be in the form of a table as follows 
\begin{align}
[s_i, V_o(s_i)] \quad i = 1, \ldots, n
\end{align}
where $s_i$ are SOC values in the increasing order such that $s_1 = 0$ and $s_n=1$ and $V_o(s_i)$ are the corresponding averaged OCV (also referred to in this paper as the pseudo-OCV). 
Typical values of $n$ can be in the thousands for low-rate OCV data that is usually recorded once every minute. 

Instead of having to store thousands of OCV-SOC pairs, typical BMSs employ curve-fitting methods to reduce the storage requirements. 
The curve fitting approaches employ functions that are based on the electrochemical behavior of the OCV described in \eqref{eq:Nernst}. 
A list of OCV-SOC models and their modeling error analysis are reported in \cite{pattipati2014open,pillai2022open} and the references therein. 

The curve fitting error at a certain SOC is then modeled as 
\begin{align}
\tilde E_{\rm CF}(s) \sim \cN (\mu_{\rm CF}(s), \sigma_{\rm CF}^2 (s) )
\end{align}
where $s$ denotes SOC, and $\tilde E_{\rm CF}(s)$ denotes the modeling error in OCV due to curve fitting at a certain SOC $s \in [0,1].$

Curve fitting errors of OCV-SOC models have been widely studied and reported in the literature \cite{pattipati2014open,pillai2022open}.
Based on the reported results, it can be concluded that the variance of the curve fitting error changes significantly with SOC. It must be noted that curve fitting error only applies if the OCV-SOC data is stored as a parametric curve. If the BMS chooses to store OCV-SOC data as (thousands of) OCV-SOC pairs, then the curve fitting error can be neglected.

\subsection{Measurement and Estimation Error}
\label{sec:MeasEstError}

Two important applications of the OCV-SOC curve are in SOC estimation and in battery capacity estimation detailed in Section \ref{sec:SOClookup} and Section \ref{sec:CapacityEstimation}, respectively.
In these two examples, it is assumed that the OCV is perfectly known. 
The uncertainty $\tilde E$ in \eqref{eq:Ehat} refers to modeling error.
However, in practice, the OCV needs to be either measured or estimated:
\begin{itemize}
\item 
Measured OCV. 
To measure the OCV, the battery must be relaxed first. This may take several hours. 
The error in this cause could be due to insufficient relaxation, hysteresis effect, and the sensitivity of the measurement device. 
Low-cost sensors can introduce significant measurement errors. 
It was argued in \cite{bfg_part1} that these errors can be considered zero-mean. 
\item
Estimated OCV. 
In most battery management systems, the OCV is estimated in real-time by modeling the voltage drop through electrical equivalent circuit models (ECMs) and employing filtering techniques, such as, 
extended Kalman filter (EKF) \cite{plett2004extended,hu2017condition,movahedi2021hysteresis,yuan2022state,shi2022state}, 
unscented Kalman filter (UKF) \cite{cui2021state,zhu2021novel}, and
cubature Kalman filter (CKF) \cite{fu2022state,ling2021state,li2022cubature,xia2017cubature}, and 
particle filter (PF) \cite{lai2021novel,khaki2021equivalent}.   
All these filtering approaches need to model the OCV estimation error \cite{balasingam2021identification,pillai2022optimizing}. 
\end{itemize}

The measurement and estimation error at a certain SOC is then modeled as 
\begin{align}
\tilde E_{\rv}(s) \sim \cN (\mu_{\rv}(s), \sigma_{\rv}^2 (s) )
\end{align}
where 
$s$ denotes SOC, and
$\tilde E_{\rv}(s)$ denotes the error in the measured/estimated OCV at a certain SOC $s \in [0,1].$
Based on results reported in the literature, it may be reasonable to assume that the measurement/estimation error is not affected by the SOC, however, all the errors are treated as functions of the SOC in this paper. 

\subsection{SOC Lookup Error}

The SOC lookup error is defined in Section \ref{sec:SOClookup} as follows
\begin{align}
\tilde s \sim \cN (\mu_\rs, \sigma_\rs^2 )
\end{align}
where the variance of the error $\sigma_s^2,$ defined in \eqref{eq:SOCerrorVar}, is shown to be a function of the uncertainty of the OCV model defined in $\sigma_E^2(s).$ 
Based on the various sources of errors defined earlier in this section, the variance of the OCV error can be written as 
\begin{align}
\sigma_\rE^2(s) &=  \sigma_{\rm c2c}^2 (s) + \sigma_{\rT}^2 (s) + \sigma_{A}^2 (\rs)  + \\ \nonumber
&\quad \quad \quad \sigma_{\rm crate}^2 (s) + \sigma_{\rm CF}^2 (s) + \sigma_{\rv}^2 (s)
\end{align}
It must be noted that the standard deviation of the SOC lookup error is dependent on the other six sources of error described in Section \ref{sec:C2Cvar} to Section \ref{sec:MeasEstError}.
In the absence of such uncertainties, the SOC can be perfectly estimated. 
That is, when
\begin{align}
 \sigma_{\rm c2c}^2 (s) =  \sigma_{\rT}^2 (s) =  \sigma_{\rA}^2 (s)=  \sigma_{\rm crate}^2 (s) = \sigma_{\rm CF}^2 (s)  = \sigma_{\rv}^2 (s) = 0
\end{align}
that will lead to $\sigma_E^2(s) = 0$ in \eqref{eq:SOCerrorVar} and the SOC estimation becomes perfect.

\section{Conclusions}
\label{sec:conc}

The state of charge of a battery is crucial information needed for the reliable operation of a battery management system (BMS). Extensive research is done on improving SOC estimation assuming that the OCV-SOC model is perfectly known.
In most works, the SOC estimation algorithm is improved but these approaches are still reliant on the OCV-SOC curve.
The effect of error in the OCV modeling was not accounted for in these SOC estimation approaches. 
Thus, the aim of this paper was to quantify uncertainties in the OCV-SOC curve and emphasize that ``\textit{It is not sufficient to solely develop the OCV-SOC model of a battery but to consider the possible uncertainty/uncertainties in this model that become pronounced later in real-world applications}".

\begin{figure*}[h!]
\begin{center}
\includegraphics[width = 1.6\columnwidth]{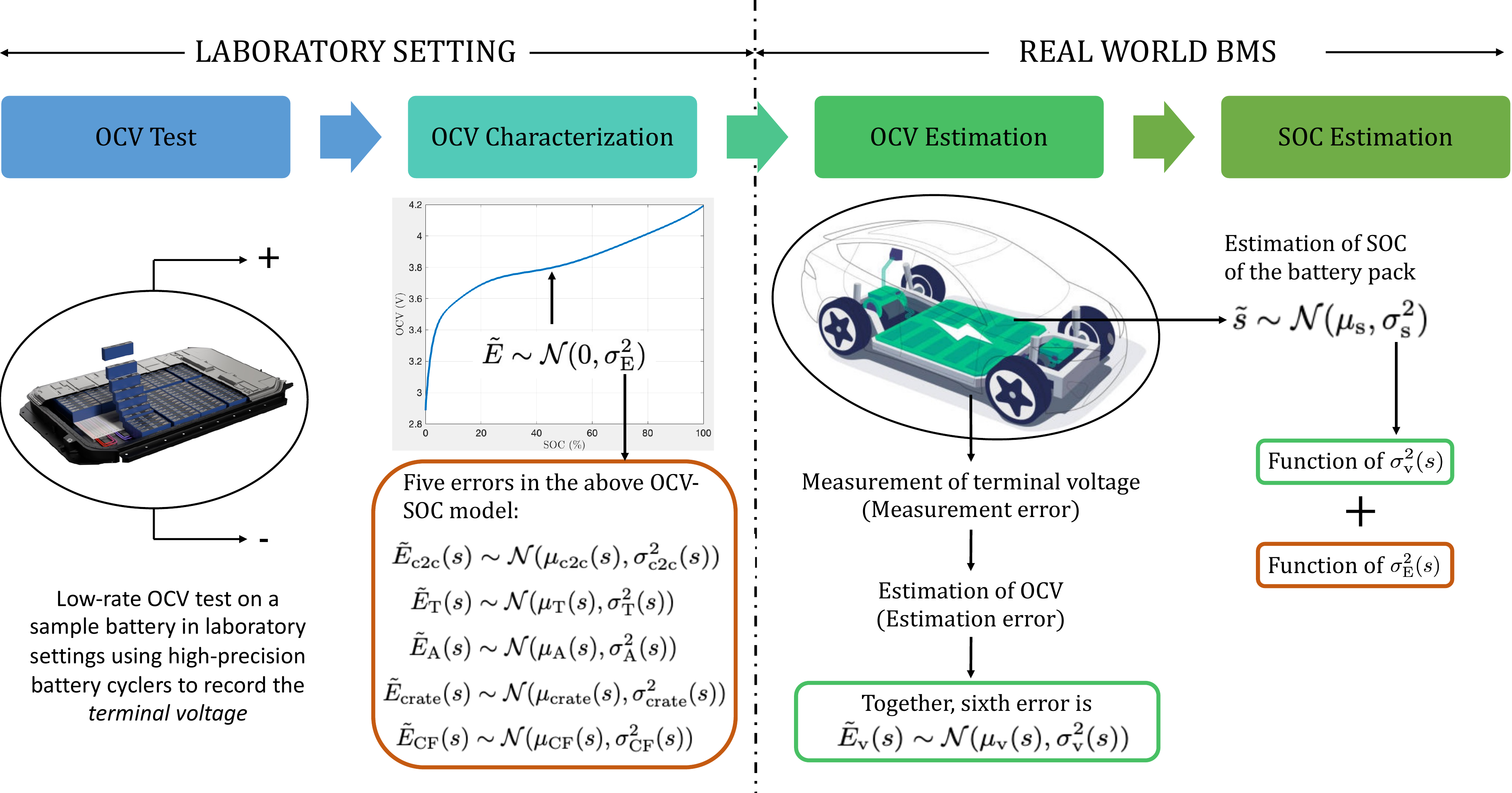}
\caption{Possible uncertainties in the OCV-SOC model and their implications in a BMS.} 
\label{fig:visconc}
\end{center}
\end{figure*}

In summary, the effect of uncertainty in the OCV-SOC curve on the SOC and battery capacity estimates is theoretically derived. 
Then, the following different sources of uncertainties to the OCV-SOC model are discussed and models are proposed to quantify them:
\begin{itemize}
\item
Cell-to-cell variation
\item
Temperature variation
\item
Aging drift
\item
Cycle-Rate error
\item
Curve Fitting error
\end{itemize}
In the BMS, the OCV-SOC curve is used to estimate the SOC (mostly in real-time). For this purpose, the measurement and estimation approaches utilized to compute the OCV are prone to errors. This results in a sixth cause of the error to be analyzed in this paper---measurement and estimation error.
Finally, the above six sources of error are combined to define a unified OCV uncertainty coefficient. Figure \ref{fig:visconc} provides a summary of this paper in graphical format.

Data collection approaches are discussed in the second part of this series to estimate the parameters of the model uncertainties presented in this paper \cite{slowOCVp2}. 
Finally, some of the model parameters are estimated and evaluated in the third part of this series of papers \cite{slowOCVp3}.

\balance

\bibliographystyle{ieeetr}
\bibliography{References,literature_BFG}

\end{document}